\documentclass[10pt,letterpaper]{article}
\usepackage{opex3}
\usepackage{setspace}
\usepackage{amsmath}
\usepackage{graphicx}
\usepackage{rangecite}

\begin{document}

\title{Precision analysis for standard deviation measurements of immobile single fluorescent molecule images}

\author{Michael C. DeSantis,$^{1}$ Shawn H. DeCenzo,$^{1}$ Je-Luen Li,$^{2}$ and Y. M. Wang$^{1,\ast}$}
\address{$^{1}$Department of Physics, Washington University, St. Louis, MO 63130, USA\\$^{2}$D. E. Shaw Research, New York, NY 10036, USA}
\email{ymwang@wustl.edu}

%
%
%




\begin{abstract}
Standard deviation measurements of intensity profiles of stationary single fluorescent molecules are useful for studying axial localization, molecular orientation, and a fluorescence imaging system's spatial resolution.  Here we report on the analysis of the precision of standard deviation measurements of intensity profiles of single fluorescent molecules imaged using an EMCCD camera.  We have developed an analytical expression for the standard deviation measurement error of a single image which is a function of the total number of detected photons, the background photon noise, and the camera pixel size.  The theoretical results agree well with the experimental, simulation, and numerical integration results.  Using this expression, we show that single-molecule standard deviation measurements offer nanometer precision for a large range of experimental parameters.
\end{abstract}




%
\section{Introduction \label{introduction}}


Single-molecule-fluorescence imaging has been a powerful tool in particle localization and tracking studies \cite{Webb2002,Selvin2003,Wang2006,Ha2008}.  In single-molecule imaging, the fluorescence intensity profile of a point light source is called a point spread function (PSF).  While the PSF is described by an Airy function, it is, in practice, approximated by a Gaussian. A Gaussian fit to the PSF of a stationary single fluorophore has two fitting parameters: centroid and standard deviation (SD).  The centroid is the center of the PSF, and the SD is its width.  While the centroid determines the lateral position of the particle in the imaging plane, the standard deviation determines its axial position \cite{Oijen1998,Florin2003,ZhuangCylinder2008} and orientation \cite{Kinosita2000,Selvin2006,Unser2009}, as well as the spatial resolution of the fluorescence imaging system \cite{Tarif2006}.    

Error analysis provides the precision for a physical measurement, and is essential for validation of the method used.  While error analysis of single-molecule PSF centroid measurements has provided the precision for lateral localization measurements \cite{Webb2002}, which enabled differentiation of various biological mechanisms (such as the walking mechanisms of myosin V on actin \cite{Selvin2003}), PSF SD measurement error analysis will provide the precision in the following applications: (1) single-molecule axial position measurements, where the SD of a single molecule's PSF increases with the defocusing distance \cite{Oijen1998,Florin2003,ZhuangCylinder2008}; (2) single-molecule orientation measurements, where at different molecular orientations with respect to the imaging plane, the molecule exhibits an elliptical-shaped Gaussian PSF with a SD that changes in both lateral directions depending on its orientation \cite{Kinosita2000,Selvin2006,Unser2009}; and (3) characterization of a single-molecule-fluorescence imaging system, where the measured SD of an imaged single fluorophore determines whether the imaging system is diffraction-limited \cite{Tarif2006}.

In contrast to the precision of centroid measurements which has been extensively investigated and applied to many systems \cite{Webb2002,Selvin2003,Selvin2004}, the precision of SD measurements of single-molecule PSFs has not been evaluated.  Here we report SD measurement error studies of immobile-single-molecule PSFs using analytical calculation, numerical integration, simulation, and experimental measurements.  As with centroid analysis, the precision of SD measurements is affected by the experimental settings of a finite number of photons per PSF $N$, the standard deviation of the background noise $\sigma_b$, and the camera's finite pixel size $a$.  We have obtained an analytical expression for the PSF SD measurement error as a function of these parameters.  Our SD measurements have achieved nanometer resolution for a wide range of experimental conditions.  This expression for the SD measurement error will provide confidence in determining a particle's axial position and molecular orientation from measurements using a single-molecule imaging system of known resolution.

\section{Theory \label{theory}}
\subsection{Formulating SD measurement error, $\Delta{s}$, by $\chi^2$ minimization}
The term ``standard deviation" was introduced by Pearson in his 1894 mathematical study of evolution \cite{Pearson1894} and characterized further in the following years \cite{Pearson1898}.  For different collections of size $N$ of randomly selected data from a common distribution with a theoretical standard deviation $s_0$, the error associated with the SD measurement of each collection is $\sqrt{s_0/2N}$, as first calculated by Pearl in 1908 \cite{Pearl1908}.  The same expression was derived more recently by Taylor using a different method \cite{Taylor1997}.  In this article, we derive the SD error for a PSF, which is a collection of photons from a common distribution emitted by a point light source.  We include the additional experimental effects of photon count fluctuation per PSF, background noise, and camera pixelation in our study. 

We utilized the method developed by Bobroff \cite{Bobroff1986} and subsequently used for centroid error analysis by Thompson, Larson, and Webb \cite{Webb2002} to derive the error associated with SD measurements of single fluorophores.  The approach uses Chi-square statistics to estimate the error associated with fitting of experimental data to expected theoretical values.  In order to maintain consistency in notation for single-molecule tracking studies, we will retain many of the same notations  used in Ref. \cite{Webb2002}.  Below, we derive the analytical solution to the PSF SD error as a function of $N$, $a$, and $\sigma_b$ ($b$ in prior studies) beginning with one dimension and extending to two dimensions.  

In 1D least squares fitting of the intensity profile of an immobile single fluorophore, $\chi^{2}(s)$ is proportional to the sum of squared errors between the observed photon count at pixel $i$, $y_i$, and the expected photon count $N_i(x, s)$, of a PSF.  Here $x$ and $s$ are the measured position and SD of the PSF, respectively, while $x_0$ and $s_0$ are the true location and the theoretical SD of the molecule:  
\begin{equation}
	\chi^{2}(s) = \sum_i\frac{(y_i-N_i)^2}{\sigma_{i,photon}^2}, 
	\label{eqn:chisq}
\end{equation}
where $\sigma_{i,photon}$ is the expected photon count uncertainty at pixel $i$ without accounting for photon-to-camera count conversion (described in the following section).  In this article, we emphasize the SD error and assume that the location measurement errors are negligible, i.e. $x = x_0$ (Appendix \ref{app2} shows that the codependence of localization and SD errors vanishes).  For simplicity, $N_i(x_0,s)$ is denoted as $N_i$ in this article unless otherwise specified.        

There are two sources for $\sigma_{i,photon}$ at pixel $i$: one is the Poisson-distributed photon shot noise of the PSF where the variance is the mean expected photon count of the pixel, $N_i$, and the other is the SD of the background noise, $\sigma_b$, expressed in photons.  The variances of the two sources add to yield
\begin{equation}
	\sigma_{i,photon}^{2} = N_i + \sigma_b^{2}.
	\label{eqn:uncerts1}
\end{equation}

The deviation of $s$ from $s_0$, $\Delta{s} = s-s_0$, is obtained by setting $d\chi^{2}(s)/ds$ to 0, expanding $N_i$ about $s_0$, and keeping the first order term in $\Delta{s}$:
\begin{eqnarray}
	\lefteqn{\Delta s}\;\;\;\;
	& = & -\frac{\sum_i\frac{\Delta y_{i}N_{i}'}{\sigma_{i,photon}^{2}}\left(1-\frac{\Delta y_{i}}{\sigma_{i,photon}^{2}}\right)}{\sum_i\frac{N_{i}'^{2}}{\sigma_{i,photon}^{2}}\left(1-\frac{2\Delta y_{i}}{\sigma_{i,photon}^{2}}\right)} \label{eqn:uncerts2a}\\
	& \approx & -\frac{\sum_i\frac{\Delta y_{i}N_{i}'}{\sigma_{i,photon}^{2}}}{\sum_i\frac{N_{i}'^{2}}{\sigma_{i,photon}^{2}}} \label{eqn:uncerts2},
\end{eqnarray}
where $N_i'$ is the derivative of $N_i$ with respect to $s$ evaluated at $s_0$, and $\Delta{y_i} = N_i(x_0,s_0)-y_i$.  By squaring Eq. (\ref{eqn:uncerts2}) we obtain the mean squared value of $\Delta{s}$,

\begin{equation}
	\langle(\Delta s)^{2}\rangle = \frac{1}{\sum_i(N_i'^2/\sigma_{i,photon}^2)}.
	\label{eqn:sum1}
\end{equation}
The root mean square of $\Delta{s}$, $\Delta{s}_{rms}$, is the PSF SD error that we calculate in this article.  Appendix \ref{app1} shows the detailed derivation of Eq. (\ref{eqn:sum1}) from $d{\chi}^2 (s)/ds = 0$.

\subsection{Modifying $\sigma_{i,photon}$ to include camera count conversion effects}
When an EMCCD (Electron Multiplying Charge Coupled Device) camera is used in imaging single fluorescent molecules, the detected pixel reading is in camera counts.  In converting from camera counts to photon counts, an additional variance in $\sigma_{i,photon}$ appears.  Below we derive the uncertainty in photon counts, $\sigma_i$, to use in place of $\sigma_{i,photon}$ in Eq. (\ref{eqn:sum1}) for experiments where EMCCD camera count conversions are involved.  

An EMCCD camera amplifies the detected photons by converting each photon to a distribution of photoelectrons through many multiplication stages.  At the final stage, one photon yields a distribution of camera counts (equivalent to the last stage photoelectron counts) with a distribution function $f(n^{\ast})$ \cite{Ulbrich2007},
\begin{equation}
	f(n^{\ast}) = \frac{1}{M}\exp{\left(-n^{\ast}/M\right)},
	\label{PhotontoPhotoelectron}
\end{equation}
where $n^{\ast}$ is the camera counts in the distribution and $M$ is the photon multiplication factor of the camera.  Here we use $^{\ast}$ to denote camera counts in order to differentiate from photon counts.  The $n^{\ast}$ distribution has a mean of $M$ and a variance of $M^2$.  

At pixel $i$, the PSF photon count distribution is described by a Poisson distribution with the variance being equal to the mean.  Each photon at the pixel contributes two terms to the pixel's camera count variance: the mean photon shot noise variance ${M^2}$ (variance of a single photon count, which is one, multiplied by the square of the multiplication factor), and the photon-to-camera count conversion variance $M^2$.  The total camera count variance contributed by one photon is $2M^2$; therefore, a mean of $N_i$ photons yields a camera count variance of $2N_i{M^2}$.  This variance agrees with the expression in Ref. \cite{Nishiwaki2003} where the variance in camera counts $\sigma_{out,camera}^2$, is related to the variance in photon counts $\sigma_{in,photon}^2$, by an excess noise factor $F^2$, 
\begin{equation}
F^2=\frac{1}{M^2}\frac{\sigma_{out,camera}^2}{\sigma_{in,photon}^2}\approx 2
\end{equation}
for EMCCD cameras with a large number of multiplication stages.  

Fluorescence from buffer, diffusing molecules in the solution, and camera counts from electronic readout and thermal noise constitutes the total background photon count at pixel $i$, with a variance of $\sigma_b^2$ and a mean of $\langle{b}\rangle$.  The total background variance in camera counts is the sum of the background count variance $\sigma_b^2M^2$, and the variance introduced by the average number of background photons, $\langle{b}\rangle$, each with a variance of $M^2$: $(\sigma_b^{2} +\langle{b}\rangle)M^2$.

Summing the PSF and the background contributions, the total camera count variance at pixel $i$ is
\begin{equation}
	\sigma_i^{\ast{2}} = 2N_i{M^2} + (\sigma_b^{2} +\langle{b}\rangle)M^2.
	\label{eqn:camerauncerts}
\end{equation}
When expressed in photon counts, 
\begin{equation}
	\sigma_i^2 = \sigma_i^{\ast{2}} /M^2 = 2N_i + \sigma_b^2 +\langle{b}\rangle.
	\label{eqn:photonuncerts}
\end{equation}  
Revising Eq. (\ref{eqn:sum1}) with the modified $\sigma_i$ we have 
\begin{equation}
	\langle(\Delta s)^{2}\rangle = \frac{1}{\sum_i(N_i'^2/\sigma_i^2)}.
	\label{eqn:pesum}
\end{equation}

\subsection{Expressing $\Delta{s}$ in photon counts}
To evaluate Eq. (\ref{eqn:pesum}) in 1D, we use a normalized Gaussian distribution
\begin{equation}
	N_{i} = \frac{Na}{\sqrt{2\pi}s}\exp{\left(-(ia)^2/2s^2\right)},
	\label{eqn:N2}
\end{equation}
where we set the location of the PSF to be at $x_0=0$ for simplicity and without loss in generality.  We approximate the pixel summation in Eq. (\ref{eqn:pesum}) by an integral going from negative to positive infinity, and we estimate $\langle(\Delta s)^{2}\rangle$ at the two extrema of $\sigma_i^2$: the high photon count regime where $\sigma_b^2 + \langle{b}\rangle$ can be neglected, and the high background noise regime where $2N_i$ can be neglected.  In the high photon count regime,
\begin{equation}
	\langle(\Delta s)^2\rangle = \frac{s_0^2}{N},
	\label{eqn:1Dphotonnoise}
\end{equation}
and in the high background noise regime, 
\begin{equation}
	\langle(\Delta s)^2\rangle = \frac{8\sqrt{\pi}{s_0}^3(\sigma_b^2+\langle{b}\rangle)}{3aN^2}.  
	\label{eqn:1Dbgnoise}
\end{equation}
An alternative derivation of Eq. (\ref{eqn:1Dphotonnoise}) is presented in Ref. \cite{Taylor1997}, although the photon-to-camera count conversion variance was not included and thus $\langle(\Delta s)^2\rangle = s_0^2/2N$.  The total 1D $\langle(\Delta s)^{2}\rangle$ is the sum of Eqs. (\ref{eqn:1Dphotonnoise}) and (\ref{eqn:1Dbgnoise}) (without the pixelation effect discussed below)
\begin{equation}
	\langle(\Delta s)^2\rangle = \frac{s_0^2}{N} + \frac{8\sqrt{\pi}s_0^3(\sigma_b^2+\langle{b}\rangle)}{3aN^2}.
	\label{eqn:1DSDerrorNoPixel}
\end{equation}

The method of approximating $\langle(\Delta s)^{2}\rangle$ by summing these results for both extrema of $\sigma_i^2$ is validated by numerical calculation results shown in Fig. \ref{Fig2}, and is in accordance with Ref. \cite{Webb2002}.  We now calculate the effect of camera pixelation on $\langle(\Delta s)^2\rangle$.  Each photon in a PSF is associated with two variances with respect to the centroid.  One is the mean variance of the PSF, $s_0^2$, and the other is due to the fact that each photon is further binned into a pixel that has an intensity profile described by a uniform distribution with a width corresponding to the pixel size $a$.  The variance of this distribution is $a^2/12$.  Thus, the total variance  of a photon due to pixelation is the sum of the two,
\begin{equation}
	s_0^{\dagger{2}} = s_0^2+\frac{a^2}{12}.
	\label{eqn:tophat}
\end{equation}
Under experimental conditions, the measured $s$ should be $\left(s_0^2+a^2/12\right)^{1/2}$ and for theoretical formulations, the expected SD of a PSF should include the pixelation effect.  We have verified that $s_0^{\dagger{2}}$ increases with $a$ according to Eq. (\ref{eqn:tophat}) by simulation. Plugging Eq. (\ref{eqn:tophat}) into Eq. (\ref{eqn:1DSDerrorNoPixel}) we have for 1D
\begin{equation}
	\langle(\Delta s)^2\rangle = \frac{s_0^2 + \frac{a^2}{12}}{N} + \frac{8\sqrt{\pi}(s_0^2+\frac{a^2}{12})^{3/2}(\sigma_b^2+\langle{b}\rangle)}{3aN^2}.
	\label{eqn:1DSDerror}
\end{equation}

Extending the 1D $\langle(\Delta s)^2\rangle$ calculation to 2D where $s_{x,y}$, which for the remainder of this article, represents the SD in either the $x$ or $y$ direction of the imaging plane, and $s_{0x}$ and $s_{0y}$ are the theoretical SD values in the $x$ and $y$ directions, respectively,

\begin{equation}
	\langle(\Delta s_x)^2\rangle = \frac{s_{0x}^2 + \frac{a^2}{12}}{N} + \frac{16\pi (s_{0x}^2 + \frac{a^2}{12})^{3/2}(s_{0y}^2+\frac{a^2}{12})^{1/2}(\sigma_b^2+\langle{b}\rangle)}{3a^2N^2}.
	\label{eqn:sigma2dp3}
\end{equation}
The derivation of Eq. (\ref{eqn:sigma2dp3}) is provided in Appendix \ref{app3}.

A more accurate estimation of $\langle(\Delta s_{x,y})^2\rangle$ can be obtained by numerically integrating Eq. (\ref{eqn:pesum}),  incorporating the transition region between the high photon  count and the high background noise regimes.  The numerical integration results are shown in Fig. \ref{Fig2} to be consistently higher than the analytical calculation results by $\approx 15 \%$.

\section{Methods \label{methods}}
\subsection{Experimental setup}
Single-molecule imaging was performed using a Nikon \textit{Eclipse} TE2000-S inverted microscope (Nikon, Melville, NY) attached to an iXon back-illuminated EMCCD camera (DV897ECS-BV, Andor Technology, Belfast, Northern Ireland).  Prism-type Total Internal Reflection Fluorescence (TIRF) microscopy was used to excite the fluorophores with a linearly polarized 532 nm laser line (I70C-SPECTRUM Argon/Krypton laser, Coherent Inc., Santa Clara, CA) focused to a 40 $\mu$m $\times$ 20 $\mu$m region on fused-silica surfaces (Hoya Corporation USA, San Jose, CA).  The incident angle at the fused-silica water interface was $64^{\circ}$ with respect to the normal.  The laser was pulsed with illumination intervals between 1 ms and 500 ms and excitation intensity between 0.3 kW/cm$^2$ and 2.6 kW/cm$^2$.  By combining laser power and pulsing interval variations we obtained 50 to 3000 photons per PSF.   A Nikon 100X TIRF objective (Nikon, 1.45 NA, oil immersion) was used in combination with a 2X expansion lens, giving a pixel size of 79 nm.  

At focus, the PSF image generated by a point light source with a mean emission wavelength of 580 nm and symmetric polarization has a full width at half-maximum (FWHM) of $\approx \lambda/2$NA $= 580$ nm$/2.9 \approx 200$ nm and theoretical $s_0 =$ FWHM/2.35 $\approx 85$ nm.  Including the pixelation effect [Eq. (\ref{eqn:tophat})], the measured PSF SD $s_{0x,0y}^{\dagger}$, for our imaging system should be 88 nm.  Due to random fluctuations in the emission polarization direction of streptavidin-Cy3 molecules attached to surfaces \cite{Kinosita2000} and variations in focus between each measurement, we observed a range of $s_{0x,0y}^{\dagger}$ values from 90 nm to 140 nm.  

Single streptavidin-Cy3 molecules  (SA1010, Invitrogen, Carlsbad, CA; 530/10 excitation, 580/60 emission) were immobilized on fused-silica surfaces by depositing 6 $\mu$l of 0.04 nM streptavidin-Cy3 powder dissolved in 0.5X TBE buffer (45 mM Tris, 45 mM Boric Acid, 1 mM EDTA, pH 7.0).  A coverslip flattened the droplet and its edges were sealed with nail polish.   The fused-silica chips were cleaned using oxygen plasma before use.  We inspected for possible surface fluorescence contaminations by imaging the TBE buffer alone; no impurities were found on either the fused-silica surface or in the buffer.  The immobilization of the adsorbed molecules was verified by centroid \textrm{vs} time measurements.

\subsection{Data acquisition and selection}
Typical movies were obtained by synchronizing the onset of camera exposure with laser illumination for different intervals.  The gain levels of the camera were adjusted such that none of the pixels of a PSF reached the saturation level of the camera.  For the initial step, streptavidin-Cy3 monomers were first selected in \textsc{ImageJ} (NIH, Bethesda, MD) by examining the fluorescence time traces of the molecules for a single bleaching step \cite{Wang2005}.  For a selected monomer, the intensity values for $25 \times 25$ pixels centered at the molecule were recorded.  The center $15 \times 15$ pixels of the PSF were used for 2D Gaussian fitting with peripheral pixels used for background analysis.  

The intensity values of the selected molecules were first converted to photon counts (see the following section) and then fitted to the following 2D Gaussian function using a least squares curve fitting algorithm (lsqcurvefit) provided by \textsc{MATLAB} (The Mathworks, Natick, MA):  
\begin{equation}
	f(x,y)=f_{0}\exp{\left(-\frac{(x-x_0)^2}{2s_x^2} - \frac{(y-y_0)^2}{2s_y^2}\right)} + \langle{b}\rangle,
	\label{eqn:gauss2dfxn}
\end{equation}
where $f_0$ was the amplitude and $\langle{b}\rangle$ was the mean background value.  A background pixel's total count is the sum of the floor, electronic readout noise, and background fluorescence counts.  For the $\langle{b}\rangle$ in this article, the floor value, determined by the lowest background pixel value, has already been subtracted.  With this fitting, the PSF's SD values in both the $x$ and $y$ directions, its measured location ($x_{0}, y_{0}$), and the image's mean background value were obtained.

The selected streptavidin-Cy3 monomers were further characterized to satisfy the following conditions used for SD error analysis.  (1) No stage drift detected by using centroid \textrm{vs} time measurements.  Stage drift introduces additional blur to each single-molecule PSF and thus affects the measured SD values.  (2) A minimum of 75 valid PSF images, each with a photon count $N$ that fluctuated less than $20\%$ from the experimental mean $\langle{N}\rangle$, of the monomer.  The PSF $N$ count restriction is necessary for precise SD error analysis at $N$ by using a statistically sufficient number of PSFs with consistent $N$.  (3) PSFs with signal-to-noise ratios ($I_0/\sqrt{I_0+\sigma_b^2}$) larger than 2.5, where $I_0$ is the peak PSF photon count (total photon count minus $\langle{b}\rangle$) and $\sigma_b^2$ is the background variance in photons.  (4) Mean $\langle{s_x}\rangle$ and $\langle{s_y}\rangle$ obtained by Gaussian fitting of the $s_x$ and $s_y$ distributions of all valid images did not differ by more than 10 nm, or $\pm 5 \%$ of the mean SD value to minimize polarization effects of Cy3.  (5) The mean SD values $\langle{s_{x,y}}\rangle$ were between 95 nm and 135 nm to minimize defocusing effects.  These constraints on $s_x$ and $s_y$ are necessary for obtaining the expression for $\Delta{s_{rms}}$, as a function of $N$, with minimal variations in the other parameters.

\subsection{Photon gain calibration}
To convert from a pixel's camera count to photons, the camera count value was divided by $M$.   In order to obtain $M$ for each experimental setting, the center nine pixel values of the PSF were evaluated if the molecule's average signal-to-noise ratio was greater than 3.  Conversely, when the signal-to-noise ratio was less than 3, only the center pixel was used for calculation as the adjacent pixels contained too few photons for statistically accurate calculations.  According to Eq. (\ref{eqn:photonuncerts}),   
\begin{equation}
M=(\sigma_{i}^{\ast2}-\sigma_b^{\ast2})/2(\langle{N_{i}^{\ast}}\rangle-\langle{b^{\ast}}\rangle),
	\label{eqn:photonConv}
\end{equation}
where $\langle{N_i}^{\ast}\rangle$ and $\sigma_{i}^{\ast}$ are the Gaussian fitted mean and standard deviation of the measured camera count distribution of pixel $i$, respectively.  Here $\langle{N_i^{\ast}}\rangle$ is the mean camera count that includes background fluorescence and electronic noise counts. For each image, the average $M$ for all nine center pixels of the PSF (or just the center pixel for PSFs with low signal-to-noise ratios) was obtained. Then, the average $M$ values for all fitted PSFs in a movie were again averaged to obtain the camera's multiplication factor for this movie.  

In order to verify that our method of calculating $M$ is correct, we have simulated PSFs with low and high $N$, including background photon noise and confirmed that $N_i$ follows a Poisson distribution which approaches a Gaussian at high $N$ with variance $N_i + \sigma_b^2$.  Including the photon-to-camera count conversion variance [Eq. (\ref{PhotontoPhotoelectron})] in the simulation, we verified Eq. (\ref{eqn:photonConv}).

\subsection{PSF and background simulations}
\begin{figure}[htbp]
	\centering\includegraphics[width=3.37in]{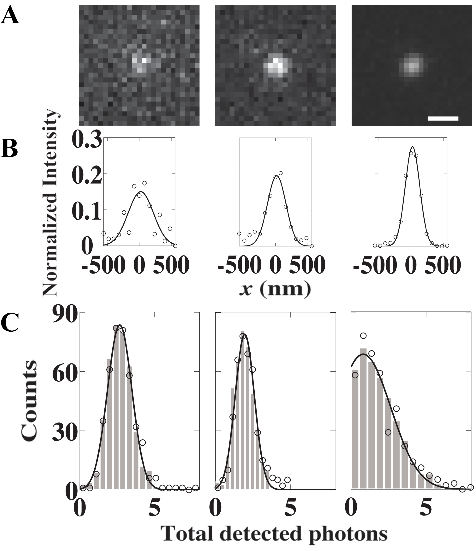}
	\caption{\label{Fig1} (A) Representative images with increasing $N$ of 151, 393, and 1891 photons of single streptavidin-Cy3 molecules.  It is evident that the ``blurriness'' of the molecules decreases with increasing $N$.  The ``blurriness" is defined to be the measured SD of the image and is the sum of the mean SD of the molecule's PSF and its SD error.  (B) 1D intensity profiles (circles) of the molecules in (A) and their Gaussian fits (lines).  The respective 1D SD values are  195.4 nm, 140.5 nm, and 110.9 nm, and the respective deviations of the 2D SD values of the images from their means are 10.3 nm, 7.2 nm, and 2.7 nm.  As expected, these deviations from the mean decrease with increasing $N$.  The scale bar is 500 nm. (C) Background count distributions (circles) for the three molecules in (A) and their fits (lines).  The histograms are simulated background distributions which reproduce those observed experimentally.}
\end{figure}

Single-fluorescent-molecule PSFs were generated using the Gaussian random number generator in \textsc{MATLAB}.  For Fig. \ref{Fig2}, the $s_{0x,0y}$ of each simulated PSF was determined by the experimental means $\langle{s_{x,y}}\rangle$.  The observed fluctuation in the number of photons $N$, was incorporated.  The generated photons of each PSF were binned into $15 \times 15$ pixels with a pixel size of 79 nm.  Then each photon count in a pixel was converted into camera count using Eq. \ref{PhotontoPhotoelectron} with a $M$ value of one.  Random background photons at each pixel were generated using the corresponding experimental background distribution function.  Although the exact experimental background distributions were used for the simulations,  the numerical integrations and analytical calculations were computed using the theoretical variance and the mean of all background counts, $\sigma_b^2$ and $\langle{b}\rangle$, respectively, rather than their fitted values.  The background counts are primarily drawn from two types of distributions: a full Gaussian with a high mean or a truncated Gaussian with a low mean (Fig. \ref{Fig1}C), depending on the background fluorescence level of each specific experiment.  The final simulated PSFs with background noise were fitted to a 2D Gaussian [Eq. (\ref{eqn:gauss2dfxn})] to obtain the centroid and SD values of the PSF.


For each simulated $\Delta{s_{x,y,rms}}$ data point, 1000 iterations (2000 iterations for Fig. \ref{Fig3}) were performed and the Gaussian fitted SDs of the $s_{x,y}$ distributions were the simulated $\Delta{s_{x,y,rms}}$ results.

\section{Results \label{results}}
We report our study of 2D $\Delta{s_{x,rms}}$ using four different methods:  (1) experimental measurements, (2) simulations, (3) numerical integrations of Eq. (\ref{eqn:pesum}), and (4) analytical calculations using Eq. (\ref{eqn:sigma2dp3}).  

Figure \ref{Fig1}A shows a set of single streptavidin-Cy3 molecule images with an increasing number of detected photons $N$.  These molecules have similar mean SD $\langle{s_x}\rangle$ values of 110 nm, 111 nm, and 107 nm, respectively.  In order to demonstrate the decreasing SD error with increasing $N$, each representative image was chosen such that the 2D SD value was the sum of the mean SD $\langle{s_x}\rangle$, and one standard deviation of the molecule's $s_x$ distribution $\Delta{s_{x,rms}}$ (SD$_{image} = \langle{s_x}\rangle + \Delta{s_{x,rms}}$).  To clearly illustrate the change in the SD error, which is measured as the PSF SD minus $\langle{s_x}\rangle$, the 1D intensity profiles of the PSFs are plotted in Fig. \ref{Fig1}B as opposed to their 2D intensity profiles for clarity.  The 1D intensity values were obtained by averaging transverse pixel intensity values of the PSF at each longitudinal pixel $i$.   It is evident that the widths of the 1D Gaussian fits decrease with increasing $N$.  The measured 2D SD$_{image}$ values deviate from their respective means, $\langle{s_x}\rangle$ values, by 10.3 nm, 7.2 nm, and 2.7 nm.  Again as expected, when $N$ increases, the 2D SD error decreases.

Figure \ref{Fig1}C presents the background distributions associated with the molecules in Fig. \ref{Fig1}A.  The background distribution function resembles either a full or a truncated Gaussian, depending on experimental settings.  The corresponding $\sigma_b$ and $\langle{b}\rangle$ values for these images are 0.92 and 2.75 photons, 0.81 and 2.07 photons, and 1.48 and 1.97 photons, respectively.  The lines are fits to the distribution and the histograms represent our simulated results.

\begin{figure}[htbp]
	\centering\includegraphics[width=3.37in]{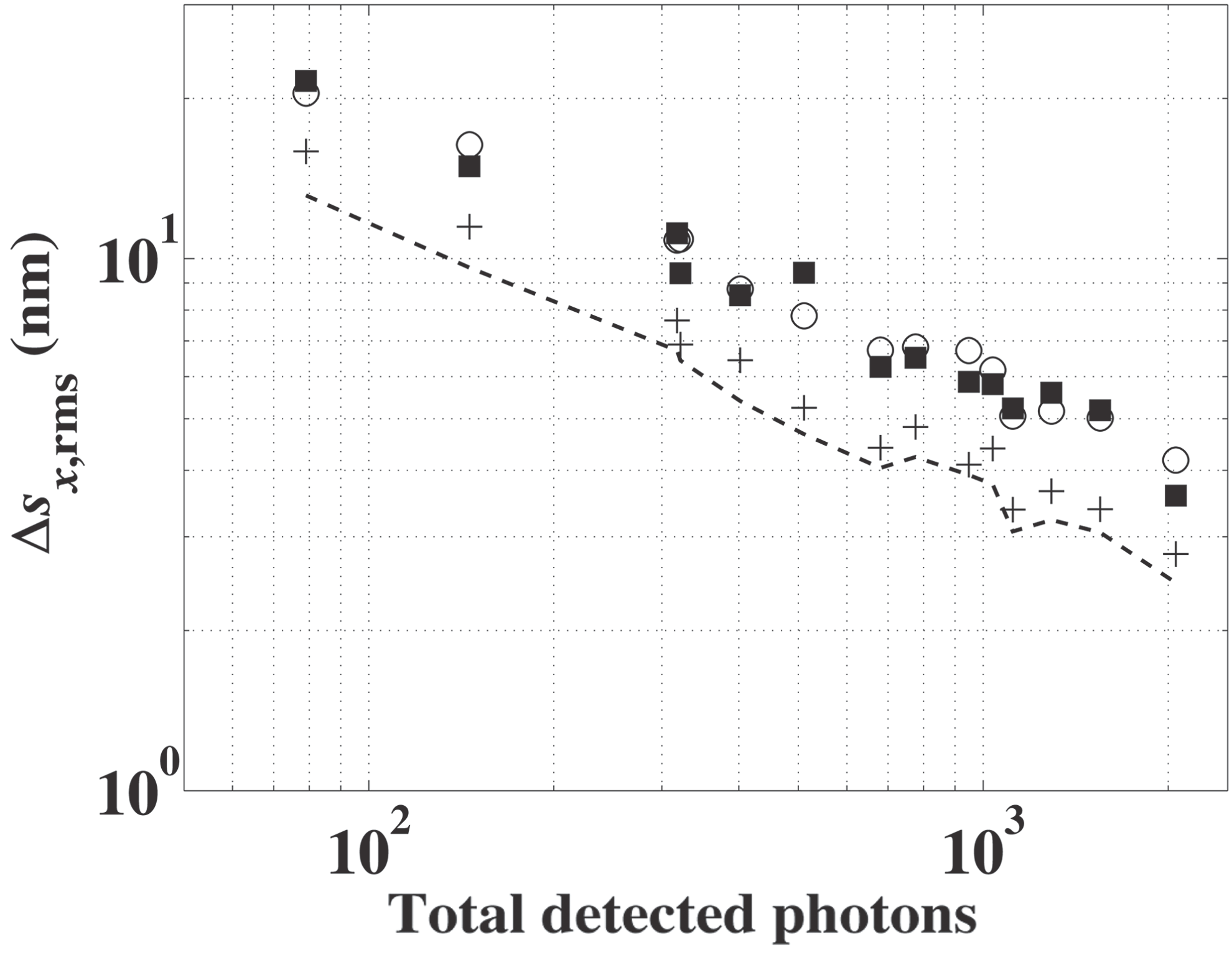}
	\caption{\label{Fig2} Comparing $\Delta{s_{x,rms}}$ \textrm{vs} $N$ obtained by using four different methods: experimental measurements (solid squares), simulations (circles), numerical integrations (crosses), and analytical calculations (dashed line).  Each experimental $\Delta{s_{x,rms}}$ data point is the SD from the Gaussian fit to the $s_x$ distribution of a single streptavidin-Cy3 monomer.  For each data point, its experimental $N$ and background distributions were used for simulation, and its experimental $\langle{N}\rangle$, $\langle{s_{x,y}}\rangle$, $\sigma_b$, and $\langle{b}\rangle$ values were used for the numerical integrations and analytical calculations.   The experimental data are on average $57\%$ higher than the analytical calculation data.}
\end{figure}

Figure \ref{Fig2} shows $\Delta{s_{x,rms}}$ obtained by using experimental measurements, simulations, numerical integrations, and analytical calculations.  Each experimental $\Delta{s_{x,rms}}$ data point is the standard deviation of the $s_x$ distribution for a single streptavidin-Cy3 monomer.  A simulation was performed for each experimental data point.  The parameters were based upon experimental results including fluctuations in a PSF's total detected photons, background distribution, and the $s_{0x,0y}$ values determined by the mean experimental $\langle{s_{x,y}}\rangle$ after subtracting for the pixelation effect [Eq. (\ref{eqn:tophat})].  The finite bandwidth of the emission filter was also taken into consideration by simulating each photon as being drawn from a PSF whose width is varied according to a Gaussian distribution centered about $s_{0x,0y}$ (with SD of 2 nm).  Numerical integrations and analytical calculations used the same $\langle{N}\rangle$, $s_{0x,0y}$, $\sigma_b$, and $\langle{b}\rangle$ as those in the corresponding experimental data point.  For all $N$, the numerically integrated $\Delta{s_{x,rms}}$ results are $\approx 15 \%$  higher than the theoretical results while the experimental results are $\approx 57 \%$ higher, and the simulations agree well with the experimental results.  

\begin{figure}[htbp]
	\centering\includegraphics[width=3.37in]{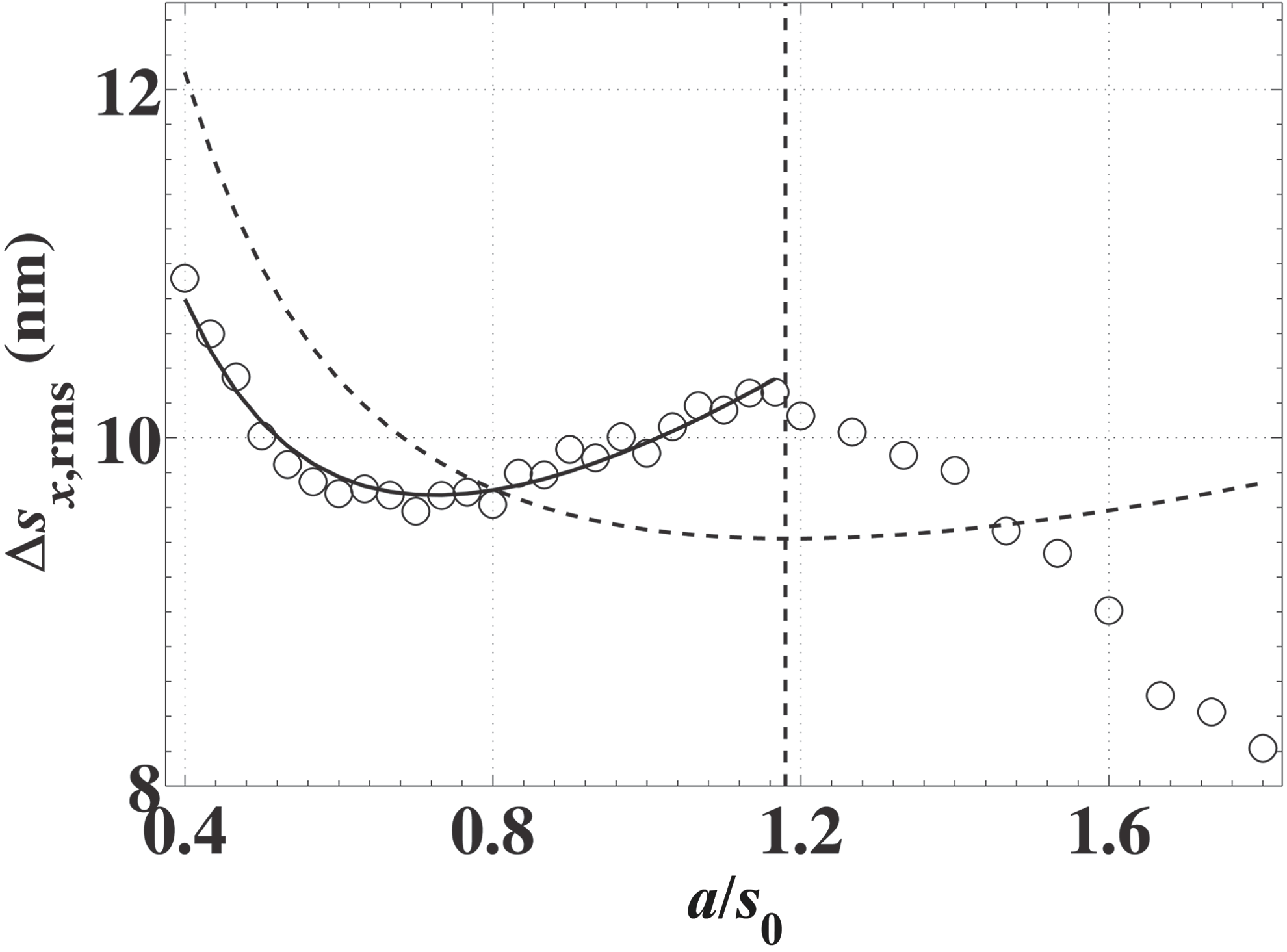}
	\caption{\label{Fig3} $\Delta{s_{x,rms}}$ \textrm{vs} $a/s_0$ studied by simulations (circles) and analytical calculations (dashed line; theoretical results shifted up by $57 \%$) for $N = 500$ photons, $s_{0} = 120$ nm, $\sigma_b = 1$ photon, and $\langle{b}\rangle = 4$ photons.  In these simulations, there were no fluctuations in $N$ and $s_{0x} = s_{0y} = s_0$.  Note that the general trend illustrates that $\Delta{s_{x,rms}}$ decreases with increasing pixel size.  For $a/s_0 < 1.18$, the simulated $\Delta{s_{x,rms}}$ results agree with the theoretical results multiplied by ($1.23 + 0.42a/s_0$) (solid line).  The shifted theory line (up by $57 \%$) crosses the simulation line at $a/s_0 \approx 0.80$ ($a \approx 96$ nm).  Considering the random discrepancy of up to $\pm 15 \%$ between simulation and experimental results in Fig. \ref{Fig2}, the cross point lies within an acceptable range given our experimental pixel size of 79 nm.  At $a/s_0 > 1.18$, the experimental $\Delta{s_{x,rms}}$ results continue to decrease influenced by the increasing dominance of the pixel's SD.  The vertical dashed line at $a/s_0 = 1.18$ is where the theoretical $\Delta{s_{x,rms}}$ minimum occurs, determined by differentiating Eq. (\ref{eqn:sigma2dp3}) with respect to $a$.}
\end{figure}

The above results are for our pixel size of 79 nm.  For different experimental settings the pixel size will vary and affect $\Delta{s_{x,rms}}$.  Figure \ref{Fig3} shows $\Delta{s_{x,rms}}$ \textrm{vs} $a/s_0$ studied by simulations and analytical calculations using $s_{0x}=s_{0y}=s_0 = 120$ nm, $N = 500$ photons, $\sigma_b = 1$ photon, and $\langle{b}\rangle = 4$ photons.  The generated photons of each PSF were binned into $19 \times 19$ pixels and subsequently converted into camera counts following the same procedure described above for Fig. \ref{Fig2}.  As $a/s_0$ increases, there is an initial decline in $\Delta{s_{x,rms}}$ until rising at $a/s_0 \approx 0.73$.  Beyond $a/s_0 \approx 0.73$, $\Delta{s_{x,rms}}$ increases slightly and then continues the decline again at $a/s_0 \approx 1.18$.  This decline after $a/s_0 \approx 1.18$ disagrees with theory, which suggests an increase in $\Delta{s_{x,rms}}$ beyond the theoretical minimum of $(a/s_0)^4 = \frac{144}{\frac{9N}{4\pi(\sigma_b^2 + \langle{b}\rangle)} + 1}$ at $a/s_0 = 1.18$ (vertical dashed line).  The overall decreasing $\Delta{s_{x,rms}}$ trend after the theoretical minimum occurs because when the pixel size increases, the measured PSF SD is increasingly affected by the width of the pixel and approaches the SD of the pixel; thus, variations among measured SD values decrease.  Eventually, at sufficiently large pixel sizes where the whole PSF is contained within one pixel, the measured SD will be the SD of the pixel, inferred by the top-hat distribution function, and the measured SD error will be zero.  The analytical calculation does not take this large pixelation effect into consideration; consequently, these results and those of the simulations begin to rapidly diverge.  

The simulated local $\Delta{s_{x,rms}}$ minimum occurs at $a/s_0 = 0.73$, rather than at the theoretical minimum of $a/s_0 = 1.18$ due to the pixel size effect described above. Our experimental settings of $a = 79$ nm and $s_0 = 120$ nm yield $a/s_0 = 0.66$ and is close to the simulated $\Delta{s_{x,rms}}$ minimum. The dashed line is the theoretical results shifted up by $57 \%$ and the solid line is the theoretical results multiplied by ($1.23 + 0.42a/s_0$), yielding an excellent fit to the simulation results for $a/s_0 < 1.18$.  We have also performed additional simulations using different parameter sets where the theoretical minimum always preceded the continued decline in $\Delta{s_{x,rms}}$.  According to Fig. \ref{Fig3} and our other simulations, a good $a/s_0$ range for future studies should be between $\approx 0.5$ and 1, as is usually the case.  Future $\Delta{s_{x,y,rms}}$ studies using different pixel sizes should take this discrepancy into account.

Note that the simulated $\Delta{s_{x,rms}}$ minimum at $a/s_0 = 0.73$ is different from the theoretical $\Delta{x_{rms}}$ minimum at $a/s_0 = 0.88$ described in Ref. \cite{Webb2002}, and our theoretical $\Delta{x_{rms}}$ minimum at $a/s_0 = 1.10$ calculated from Eq. \ref{eqn:centroiderror}, using our set of parameter values.  Future studies should take this difference into consideration by selecting an optimal pixel size.

\section{Discussion and Extensions \label{discussion}}
Here we discuss four issues: (1) causes for discrepancies between results obtained using different methods; (2) modifications to the centroid measurement error developed by Thompson, Larson, and Webb \cite{Webb2002} to include the EMCCD camera photon conversion effects; (3) relation between SD error and the error of the measured quantities associated with each of the aforementioned applications; and (4) methods to determine the SD error $\Delta{s_{x,y,rms}}$, for dimeric fluorophores and mobile molecules in future studies.

\subsection{Causes for discrepancies} 
Numerical integration results are consistently higher than the analytical results by $15 \%$, while simulation results are higher than analytical results by $57 \%$ for all $N$.  There are a number of reasons for these discrepancies: (1) The analytical $\Delta{s_{x,y,rms}}$ result [Eq. (\ref{eqn:sigma2dp3})] is obtained by evaluating Eq. (\ref{eqn:pesum}) for the two limiting cases of $\sigma_i^2$  at the high photon count and high background noise regimes.  The intermediate regime is absent and thus the numerical integration and simulation results are larger.  (2) When $N_i$ is expanded about $s_0$, the higher order terms were neglected [Eq. (\ref{eqn:uncerts2})].  (3) In the $\Delta{s_{rms}}$ calculation (Appendix \ref{app1}), the $N_i$ distribution function is assumed to be a Gaussian for all pixels of the PSF [Eq. (\ref{NiDistribution})].  This assumption will only be statistically accurate for center pixels of PSFs with high $N$.   For peripheral pixels, especially for PSFs with low $N$, the $N_i$ distribution function approaches a Poisson with a low mean, rather than a Gaussian.  These different $N_i$ distributions, which have been verified by simulation, were not considered in the analytical calculations. (4) In simulations, we attempted to model the background count distribution exactly, whereas in numerical integrations and analytical calculations, the shape of the background count distribution was not considered, and therefore did not influence the results.  

In summary, the analytical calculation of the SD measurement error expressed in Eq. (\ref{eqn:sigma2dp3}) is a reasonable approximation for a large range of experimental parameters.  When the $57 \%$ difference is corrected for, the expression is in excellent agreement with our experimental results.  Future studies using this formula should be aware of the limitations and be sure to include this $57 \%$ difference from underestimation of the true error for similar $a/s_0$ values.

\subsection{Modifications to centroid error analysis}
The PSF centroid error expression developed by Thompson, Larson, and Webb \cite{Webb2002} did not take the photon-to-camera count conversion variance into consideration.  Additionally, the theoretical standard deviation $s_0$, should be modified to include the pixelation effect $\left(s_0^2+a^2/12\right)^{1/2}$, with respect to both directions.  We have modified the PSF centroid measurement error to be

\begin{equation}
	\langle(\Delta x)^2\rangle = \frac{2(s_{0x}^2 + \frac{a^2}{12})}{N} + \frac{8\pi (s_{0x}^2 + \frac{a^2}{12})^{3/2}(s_{0y}^2+\frac{a^2}{12})^{1/2}(\sigma_b^2+\langle{b}\rangle)}{a^2N^2}.
	\label{eqn:centroiderror}
\end{equation}
This theoretical expression for the centroid measurement error underestimates the experimental results by $42 \%$.

\subsection{Interpreting $\Delta{s_{x,y,rms}}$ in SD measurement applications} 
With regards to the three applications of SD measurements discussed in this article, the SD measurement error of a single image can be translated into the precisions associated with each of the application's measured quantities. For the direct translation, SD error is the uncertainty of an imaging system's measured resolution; for the indirect translations, the precisions for axial localization and molecular orientation measurements can be expressed by SD error.

For characterization of an imaging system's spatial resolution, if the system is diffraction-limited, the SD of the imaging system should fall between the measured SD of the PSF from a visible point light source $\pm$ the SD error \cite{Tarif2006,Born1999}.  Thus, SD measurement error directly provides the precision for quantifying an imaging system's resolution. 

For axial localization studies, it has been previously shown that the SD of the PSF of a molecule located at a distance $z$ away from the focal plane can be expressed as  \cite{Oijen1998,Florin2003,ZhuangCylinder2008}
\begin{equation}
s(z) = s_0\left(1+\frac{z^2}{D^2}\right)^{1/2},
\end{equation}
where $D \approx 400$ nm is the imaging depth of a typical single-molecule imaging system.  Consequently, by error propagation, the precision in the SD measurement of a single image, $\Delta{s_{rms}}$, can be used to determine the localization error associated with the molecule's axial position, $\Delta{z}$: 
\begin{equation}
\Delta{z} = \frac{D}{s_0\left(1-\frac{s_0^2}{s(z)^2}\right)^{1/2}}\Delta{s_{rms}(z)}.
\end{equation}

For molecular orientation studies, the polarized PSF, for a range of orientations, has an elliptical intensity profile that can be fit by a 2D Gaussian with different standard deviations in the $x$ and $y$ directions \cite{Kinosita2000,Selvin2006,Unser2009}. When an expression relating $s_x$ and $s_y$ to the orientation is developed, the error in measuring $s_x$ and $s_y$, once again by error propagation, can be used to calculate an error associated with the reported orientation of the molecule.

\subsection{$\Delta{s_{x,y,rms}}$ calculation for future SD measurement applications} 
In addition to stationary single molecules, SD measurements can be used in future applications to investigate molecules such as stationary dimers or moving fluorophores.  We are currently exploring these two areas of interest: (1) two sub-diffraction limit, separated molecules labeled with identical fluorophores that exhibits a combined PSF with a SD that increases with their separation \cite{Wang2010_2}; (2) a moving molecule (i.e. a freely-diffusing fluorophore) which produces a blurred image given a finite exposure time, whereby the measured SD of the resulting intensity profile can be used to study the dynamic properties of the molecule, particularly its diffusion coefficient \cite{Wang2010_3}.  With modification, the method for estimating the SD error of stationary single molecules in this article can be extended to these two cases.  For these studies, the $N_i$ distribution function at each pixel may be different from the Gaussian assumption for stationary molecules in Eq. (\ref{eqn:N2}).   A new $N_i$ distribution function for each specific case can be obtained and a new $\sigma_i^2$ formula [Eq. (\ref{eqn:photonuncerts})] can be derived.  Using the new $N_i$ distribution function and $\sigma_i^2$, the SD error for these cases can be obtained following the same procedure outlined in the theory (Sec. \ref{theory}).

\section{Conclusion \label{conclusion}} 
In this article we report the precision analysis for SD measurements of single-fluorescent-molecule intensity profiles. Our analytical expression of the PSF SD error allows for proper quantification of the precision associated with determination of the imaging system's resolution and both axial localization and molecular orientation measurements of single molecules.  Furthermore, we propose additional studies to characterize multiple fluorophores and examine the diffusive properties of mobile molecules by evaluating the measured SDs of their corresponding intensity profile to known precision. When our theoretical framework is extended to these studies, SD analysis will be advanced into a powerful tool for single-molecule-fluorescence imaging studies.



\section*{Appendix}
\appendix
\section{\label{app1}}
Here we present the complete derivation of Eq. (\ref{eqn:sum1}).  We first obtain a probability distribution function for $y_i$. At large $N$ of a few hundred photons, the $y_i$ probability distribution function at each of the center nine pixels of the PSF is a Gaussian, while at the peripheral pixels, the $y_i$ probability distribution function is better approximated by a Poisson with a low mean.  Here we assume that our $N $ is significantly larger than 100 photons and the $y_i$ probability distribution functions for all PSF pixels are Gaussian functions 
\begin{equation}
f_{y_i} = \frac{1}{\sqrt{2\pi}\sigma_i} \exp\left(-\frac{\Delta y_i^2}{2\sigma_i^2} \right),
\label{NiDistribution}
\end{equation}
where $\Delta y_i = N_i(x_0,s_0)-y_i$ and $\sigma_i^2$ is $\sigma_{i,photon}^2$ as in Eq. (\ref{eqn:chisq}).  For Gaussian distributed $y_i$, we have
\begin{subequations}
	\begin{eqnarray}
		\langle \Delta y_i \rangle &=& 0,\\
		\langle (\Delta y_i)^2 \rangle &=& \sigma_i^2 \label{eq:3}.
	\end{eqnarray}
\end{subequations}

Starting from Eq. (\ref{eqn:chisq}) and taking a derivative with respect to $s$,

\begin{equation}
	\frac{d\chi^2 (s)}{ds} = \sum_i \frac{d}{ds}\frac{(y_i-N_i)^2}{\sigma_i^2} = \sum_i \frac{2(y_i-N_i)(y_i-N_i)'\sigma_i^2-(y_i-N_i)^2\cdot 2\sigma_i\sigma_i'}{\sigma_i^4}.
\end{equation}

Setting the above equation to zero, we find

\begin{equation}
	\sum_i \frac{2(y_i-N_i)(y_i-N_i)'}{\sigma_i^2} = \sum_i \frac{(y_i-N_i)^2\cdot 2\sigma_i\sigma_i'}{\sigma_i^4}.
	\label{eq:4}
\end{equation}
		
We can simplify Eq. (\ref{eq:4}) using the following terms:

	\begin{subequations}
		\begin{eqnarray}
		y_i - N_i(s) &=& y_i - (N_i(s_0)+N_i'\Delta s) = -\Delta y_i - N_i'\Delta s,\label{eq:5}\\
(y_i - N_i)'    &=&  -N_i',\\
\sigma_i^2     &=& 2N_i(s)+2\sigma_b^2 = 2(N_i(s_0)+N_i'\Delta s) +2\sigma_b^2, \\
2\sigma_i \sigma_i'     &=& 2N_i'. \label{eq:8}
		\end{eqnarray}
	\end{subequations}

Inserting Eqs. (\ref{eq:5})-(\ref{eq:8}) into Eq. (\ref{eq:4}), we obtain

	\begin{eqnarray}
		\sum_i \frac{-2(\Delta y_i + N_i' \Delta s)(-N_i')}{\sigma_i^2} &=& \sum_i \frac{(\Delta y_i + N_i' \Delta s)^2\cdot 2N_i'}{\sigma_i^4} \nonumber \\
&\approx& \sum_i \frac{(\Delta y_i^2 +2\Delta y_i N_i' \Delta s)\cdot 2N_i'}{\sigma_i^4}. 
		\label{eq:9}
	\end{eqnarray}

Moving $\Delta s$ to the left-hand side,
\begin{equation}
\Delta s \sum_i\left( \frac{N_i'^2 }{\sigma_i^2} - \frac{2\Delta y_i N_i'^2}{\sigma_i^4} \right)= \sum_i \left(\frac{\Delta y_i^2 N_i'}{\sigma_i^4} - \frac{\Delta y_i N_i'}{\sigma_i^2}\right). 
\end{equation}
This equation is Eq. (\ref{eqn:uncerts2a}).  Neglecting the $\Delta y_i/\sigma_i^2$ term, we get Eq. (\ref{eqn:uncerts2}).

We now take the mean square of Eq. \ref{eqn:uncerts2}.  Note that the average is meant to apply to $y_i$ only, so we have

\begin{equation}
	\langle(\Delta s)^2\rangle = \frac{\sum_i \frac{\Delta y_i N_i'}{\sigma_i^2}\sum_j \frac{\Delta y_j N_j'}{\sigma_j^2}}{\left(\sum_i \frac{N_i'^2}{\sigma_i^2}\right)^2} = \frac{\sum_{i,j} \frac{\langle \Delta y_i \Delta y_j\rangle N_i'N_j'}{\sigma_i^2\sigma_j^2}}{\left(\sum_i \frac{N_i'^2}{\sigma_i^2}\right)^2}.
\end{equation}

For two different pixels, their distributions are independent, so
$\langle \Delta y_i \Delta y_j\rangle = \delta_{ij} \langle (\Delta y_i)^2 \rangle=\sigma_i^2$ [see Eq. (\ref{eq:3})]. This gives us Eq. (\ref{eqn:sum1}).

\section{\label{app2}}
Here we calculate the codependence of $\Delta{x}$ and $\Delta{s}$ in 1D and show that it vanishes.  Thus the assumption that $x$ (or $s$) is fixed when taking a partial derivative of $N_i$ with respect to $s$ (or $x$) is valid.

Expanding $N_i(x,s)$ about $(x_0, s_0)$ to first order $=> N_i=N_i(x_0,s_0) + \Delta{x}\frac{\partial{N_i}}{\partial{x}}|_{x_0}+\Delta{s}\frac{\partial{N_i}}{\partial{s}}|_{s_0}$.  Setting $\frac{d(\chi^2)}{ds}=\frac{d(\chi^2)}{dx}=0$ and solving for   
$\Delta{s}$ and $\Delta{x}$,  the error associated with the $s$ measurement is
	\begin{equation}
\langle(\Delta{s})^2\rangle=\frac{\sum_i\frac{(\frac{\partial{N_i}}{\partial{s}}|_{s_0})^2}{\sigma_i^2}+2\frac{\left(\sum\frac{\frac{\partial{N_i}}{\partial{x}}|_{x_0}\frac{\partial{N_i}}{\partial{s}}|_{s_0}}{\sigma_i^2}\right)^2}{\frac{(\frac{\partial{N_i}}{\partial{x}}|_{x_0})^2}{\sigma_i^2}}+\frac{\left(\sum\frac{\frac{\partial{N_i}}{\partial{x}}|_{x_0}\frac{\partial{N_i}}{\partial{s}}|_{s_0}}{\sigma_i^2}\right)^2}{\sum\frac{(\frac{\partial{N_i}}{\partial{x}}|_{x_0})^2}{\sigma_i^2}}}{(\sum_i\frac{(\frac{\partial{N_i}}{\partial{s}}|_{s_0})^2}{\sigma_i^2})^2-2\left(\frac{\sum\frac{(\frac{\partial{N_i}}{\partial{s}}|_{s_0})^2}{\sigma_i^2}(\sum\frac{\frac{\partial{N_i}}{\partial{x}}|_{x_0}\frac{\partial{N_i}}{\partial{s}}|_{s_0}}{\sigma_i^2})^2}{\sum\frac{(\frac{\partial{N_i}}{\partial{x}}|_{x_0})^2}{\sigma_i^2}}\right)+\frac{(\sum\frac{\frac{\partial{N_i}}{\partial{x}}|_{x_0}\frac{\partial{N_i}}{\partial{s}}|_{s_0}}{\sigma_i^2})^4}{(\sum\frac{(\frac{\partial{N_i}}{\partial{x}}|_{x_0})^2}{\sigma_i^2})^2}}.
	\end{equation}
The cross product term $\sum_i\left(\frac{\frac{\partial{N_i}}{\partial{x}}|_{x_0}\frac{\partial{N_i}}{\partial{s}}|_{s_0}}{\sigma_i^2}\right)=0$, and arriving at Eq. (\ref{eqn:sum1}), $\langle(\Delta{s})^2\rangle = \frac{1}{\sum_i((\frac{\partial{N_i}}{\partial{s}})^2/\sigma_i^2)}$.

\section{\label{app3}}
In Appendix \ref{app3} we calculate the 2D $\langle(\Delta{s_x}^2)\rangle$.  In 2D, the expected counts at pixel $i$, $j$ is given by 
\begin{equation}
	N_{i,j}=\frac{Na^2}{2\pi{s_{x}}{s_{y}}}\exp{\left(-\frac{(ia)^2}{2s_{x}^2} - \frac{(ja)^2}{2s_{y}^2}\right)},
	\label{A3.1}
\end{equation}
where we assume that the PSF is centered at zero.  Taking the derivative of $N_i$ with respect to $s_x$ and evaluating at $s_{0x}$,  
\begin{equation}
	\langle(\Delta{s_x^2})\rangle=\frac{1}{\sum_i \frac{(\frac{d(N_i)}{d{s_x}})^2}{\sigma_i^2}}.
	\label{A3.2}
\end{equation}
Next, we approximate the summation by an integral where $i$ and $j$ are continuous from negative to positive infinity.  There are two limits to the approximation, one being the high photon count limit and the other being the high background noise limit.   At the high photon count limit, $\langle(\Delta{s_x})^2\rangle=\frac{s_{0x}^2}{N}$ after taking the photon-to-camera count conversion variance into consideration.   At the high background noise limit, $\langle(\Delta{s_x})^2\rangle=\frac{16\pi s_{0x}^3s_{0y}(\sigma_b^2+\langle{b}\rangle)}{3a^2N^2}$.  Adding the two terms together and replacing $s_{0x,0y}$ by $\left(s_{0x,0y}^2+a^2/12\right)^{1/2}$ to incorporate the pixelation effect, we arrive at Eq. (\ref{eqn:sigma2dp3}).

\section*{Acknowledgements}
We thank A. Carlsson for valuable discussions. Michael C. DeSantis is supported by a National Institutes of Health predoctoral fellowship awarded under 5T90 DA022871.

\end{document}